\documentclass[aps,prd,twocolumn,nofootinbib]{revtex4}

\usepackage{booktabs,makecell,multirow}

\usepackage{graphicx}
\usepackage{float}
\usepackage{color}

\usepackage[colorlinks=true, linkcolor=blue, citecolor=blue, urlcolor=blue]{hyperref}

\newcommand*{\RMN}[1]{\uppercase\expandafter{\romannumeral#1}}

\begin{document}

\title{Updated determination of the pion-photon transition form factor}

\author{Hua Zhou$^{1}$}
\email{zhouhua@cqu.edu.cn}

\author{Jiang Yan$^{1}$}
\email{yjiang@cqu.edu.cn}

\author{Qing Yu$^{2,1}$}
\email{yuq@swust.edu.cn}

\author{Xing-Gang Wu$^{1}$}
\email{wuxg@cqu.edu.cn}

\affiliation{$^1$ Department of Physics, Chongqing Key Laboratory for Strongly Coupled Physics, Chongqing University, Chongqing 401331, P.R. China \\
$^2$ School of Science, Southwest University of Science and Technology, Mianyang 621010, P.R. China
}

\date{\today}

\begin{abstract}

In this paper, we study the properties of the pion-photon transition form factor (TFF), $\gamma\gamma^{\ast} \rightarrow \pi^{0}$, by using the principle of maximum commonality (PMC) to deal with its perturbative QCD contribution up to next-to-next-to-leading order QCD corrections. Applying the PMC, we achieve precise pQCD prediction for the TFF in a large $Q^2$-region without conventional renormalization scale ambiguity. We also discuss the power suppressed non-valence quark contribution to the TFF, which is important for a sound prediction in low and intermediate $Q^2$-region, e.g. the non-valence quark components affect the TFF by about $1\%$ to $23\%$ when $Q^{2}$ changes down from $40~{\rm GeV^{2}}$ to $4~{\rm GeV^{2}}$. The resultant pion-photon TFF shows a better agreement with previous Belle data. It is hoped that previous discrepancies between the experimental measurements and theoretical predictions could be clarified by the forthcoming precise data on the Belle II measurements.

\end{abstract}

\maketitle

To meet the needs of the forth-coming more-and-more precise experimental measurements, it is helpful to have more precise standard model (SM) prediction so as to improve the comparison of theoretical predictions with the data and then ensure whether there is new physics beyond the SM or not. Among them, the pion-photon transition form factor (TFF) is an important component that is related to the axial anomaly~\cite{Adler:1969gk, Bardeen:1969md} and can be used to study the chiral symmetry, the quark-mass ratio, the characteristics of the pseudo-scalar meson's decay, and etc.~\cite{Cornwall:1966zz, Brodsky:1971ud}. The pion-photon TFF predicts the time-ordered product of two electromagnetic currents using the operator-product-expansion (OPE)~\cite{Lepage:1979zb, Lepage:1980fj}. Different kinematic setting functions of the TFF provide a theoretical description of various two-photon processes, e.g., the deeply inelastic lepton-hadron scattering and the deeply virtual compton scattering~\cite{Bjorken:1968dy, Bjorken:1969ja, Muller:1994ses, Radyushkin:1996nd}. The pion-photon TFF is also an important input to calculate the hadronic light-by-light contribution to the muon anomalous magnetic moment $(g-2)_{\mu}$~\cite{Colangelo:2014dfa, Hoferichter:2018kwz, Hoferichter:2018dmo}. Therefore, in-depth research on the pion-photon TFF is important.

Early experimental studies on the pion-photon TFF traced back to the CELLO and CLEO measurements in 1991 and 1998~\cite{CELLO:1990klc, CLEO:1997fho}. Lately, in 2009, the BaBar collaboration~\cite{BaBar:2009rrj} issued the data on the pion-photon TFF for the kinematic region $Q^{2}\in[4,40]~{\rm GeV}^{2}$, which sparked a heated discussion by showing the unexpected scaling violation of the TFF, e.g. its large $Q^2$ behavior contradicts the well-known asymptotic prediction which indicates that $Q^2 F_{\pi \gamma} (Q^2)$ tends to be a constant~\cite{Lepage:1980fj}. In 2012, the Belle collaboration~\cite{Belle:2012wwz} issued their measurements for the same energy region, and the data showed that the pion-photon TFF does not increase significantly in large $Q^2$ region, which has a 2$\sigma$ deviation compared with the BaBar data. Benefiting from the high luminosity and high trigger efficiency in Belle II at SuperKEKB experiment~\cite{Belle-II:2018jsg}, the total uncertainty of the system is expected to be $2$ times smaller than that of Belle I. Then it is hopeful that the above experimental discrepancy could be clarified in near future.

Theoretically, the pion-photon TFF has been studied under various approaches. Using the QCD factorization, the pion-photon TFF with one real and one virtual photon at the leading-power level (${\cal {O}}(1/Q^{2})$) can be decomposed into two parts, i.e., the perturbatively calculable coefficient function (CF) and the non-perturbative twist-two pion light-cone distribution amplitude (LCDA)~\cite{Li:1992nu, Musatov:1997pu, Li:2013xna}. Many efforts have been devoted to the calculation of CF, e.g. Refs.\cite{delAguila:1981nk, Braaten:1982yp, Kadantseva:1985kb} derived the next-to-leading order (NLO) QCD correction; Lately, the computation of next-to-next-to-leading-order (NNLO) QCD correction under the large $\beta_{0}$-approximation and the conformal scheme has been done by Refs.\cite{Melic:2001wb, Melic:2002ij}; And the complete NNLO QCD correction using the conformal symmetry~\cite{Braun:2021grd} and the hard-collinear factorization theorem~\cite{Gao:2021iqq} have been reported recently. Those works provide us with great chances to achieve precise pQCD prediction on CF.

At present, the original perturbative NNLO series of the CF still shows large renormalization scale uncertainty, and it becomes an important systematical error in their interpretation. It is noted that such large scale error is due to the mismatching of strong coupling $\alpha_s$ and its corresponding coefficients~\cite{Wu:2013ei, Wu:2014iba, Wu:2019mky}, which is unnecessary and can be removed via a proper scale-setting approach, leading to a precise prediction consistent with standard renormalization group invariance. For example, the principle of maximum commonality (PMC)~\cite{Brodsky:2011ta, Brodsky:2011ig, Brodsky:2012rj, Brodsky:2013vpa, Mojaza:2012mf} provides a systematical way to eliminate such conventional scale ambiguity by using the renormalization group equation (RGE), or the $\beta$-function. The RGE determines the correct $\alpha_s$ running behavior of the process by recursively using the $\{\beta_i\}$-terms of the perturbative series. It has been demonstrated that after applying the PMC, the resultant perturbative series becomes scale-invariant that is independent of any initial choice of renormalization scale~\cite{Wu:2018cmb}. And due to the elimination of the RGE-involved divergent renormalon terms, the convergence of the resultant CF perturbative series can be naturally improved.

In the present paper, we shall adopt the PMC single scale-setting approach~\cite{Shen:2017pdu} to achieve a scale-invariant pQCD prediction for the CF of the leading-power valence quark contribution $Q^{2} F^{\rm V}_{\pi \gamma}(Q^{2})$. By using the PMC single scale-setting approach, the residual scale dependence~\cite{Zheng:2013uja} due to the unknown even higher-order terms can be highly suppressed. There are also power suppressed non-valence (NV) quark contribution $Q^{2} F^{\rm NV}_{\pi \gamma}(Q^{2})$, and the total pion-TFF can be written as
\begin{eqnarray}
Q^{2} F_{\pi \gamma}(Q^{2}) &=& Q^{2} F^{\rm V}_{\pi \gamma}(Q^{2}) + Q^{2} F^{\rm NV}_{\pi \gamma}(Q^{2}).
\end{eqnarray}
The $Q^{2} F^{\rm V}_{\pi \gamma}(Q^{2})$ arises from the direct annihilation of $q\bar{q}$ pair in the pion into two photons, which provides the dominant contribution to the CF in the large $Q^{2}$-region. The $Q^{2} F^{\rm NV}_{\pi \gamma}(Q^{2})$ is associated with the non-perturbative high-Fock states in the pion~\cite{Radyushkin:1995pj, Huang:2006wt, Wu:2010zc, Huang:2013yya}, whose contributions can be estimated by using proper phenomenological models. And as will be shown below, they will have sizable contributions in low $Q^2$-region.

Firstly, the leading-power contribution $Q^{2} F^{\rm V}_{\pi \gamma}(Q^{2})$ can be expressed as follow~\cite{Li:1992nu, Musatov:1997pu, Li:2013xna}
\begin{eqnarray}
Q^{2} F^{\rm V}_{\pi \gamma}(Q^{2}) &=& \frac{\sqrt{2}f_{\pi}}{6}\int^{1}_{0}dx T(x,Q,\mu_{f})\phi_{\pi}(x,\mu_{f}),
\label{eq1}
\end{eqnarray}
where the pion decay constant $f_{\pi}=130.5$ MeV~\cite{ParticleDataGroup:2022pth} and the pion light-cone distribution amplitude (LCDA) $\phi_{\pi}(x,\mu_{f})$ is usually represented as a Gegenbauer polynomial expansion~\cite{Efremov:1979qk, Lepage:1980fj}, e.g.,
\begin{eqnarray}
\phi_{\pi}(x,\mu_{f})&=&6x\overline{x}\sum_{n=0,2,\cdots}a_{n}(\mu_f) C^{3/2}_{n}(2x-1),
\label{eq2}
\end{eqnarray}
where $\overline{x}=1-x$, $C^{3/2}_{n}(2x-1)$ is the Gegenbauer polynomial and $\mu_{f}$ is the factorization scale. The first moment $a_{0}$ equals to $1$ by using the normalization condition.

\begin{figure}[htb]
\centering
\includegraphics[width=0.48\textwidth]{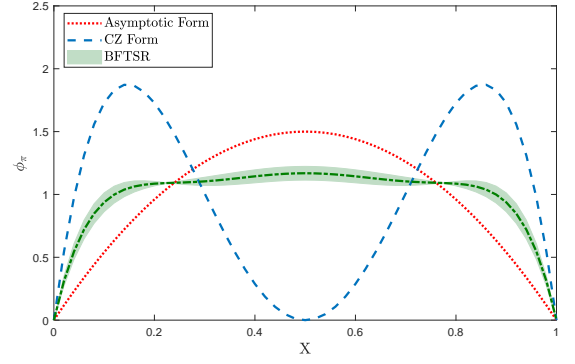}
\caption{Our present adopted LCDA at $\mu_0=1$ GeV, which is shown by the dash-dot line. The shaded band shows its uncertainty for $a_{2}=0.206\pm0.038$ and $a_{4}=0.047\pm0.011$. The dotted and the dashed are for the asymptotic LCDA and the CZ LCDA, respectively.}
\label{fig0}
\end{figure}

The pion TFF is highly sensitive to the pion LCDA~\cite{Braun:2021grd, Gao:2021iqq}. To determine the precise behavior of the pion LCDA is an interesting topic, which has been discussed under various approaches, e.g. the Lattice QCD, the QCD sum rules, the Dyson-Schwinger equations, and etc.~\cite{Brodsky:2007hb, Chang:2013pq, Raya:2015gva, Mikhailov:2016klg, Raya:2016yuj, RQCD:2019osh, Cui:2020tdf, Cheng:2020vwr, Stefanis:2020rnd, Zhong:2022lmn, Gao:2022vyh}. Most of which prefer the asymptotic form suggested by Lepage and Brodsky~\cite{Lepage:1980fj}. In this paper, we will adopt the latest Gegenbauer moments $a_{n}$ obtained by using the QCD sum rules in the framework of background field theory~\cite{Zhong:2021epq} for subsequent numerical analysis. It improves the traditional light-cone harmonic oscillator model of the pion leading-twist LCDA by introducing a new longitudinal behavior of the wave function and gives a new method for determining input parameters, see Ref.~\cite{Zhong:2021epq} for more details. Specifically, it shows that the LCDA moments, e.g., $\left< \xi^{2}\right>_{2;\pi}|_{\mu_{0}} =0.271\pm0.013$, $\left< \xi^{4}\right>_{2;\pi}|_{\mu_{0}} =0.138\pm0.010$, which lead to the Gegenbauer moments $a_{2}=0.206\pm0.038$ and $a_{4}=0.047\pm0.011$ at the scale $\mu_{0}=1$ GeV. Fig.(\ref{fig0}) shows the behaviors of our adopted LCDA $\phi_{\pi}$, which is asymptotic-like. We also present the asymptotic form ($\phi_{{\rm AS}}=6x(1-x)$ with $a_{2}=0$)~\cite{Lepage:1980fj} and the CZ form ($\phi_{{\rm CZ}}=30x(1-x)(2x-1)^{2}$ with $a_2=0.6$)~\cite{Chernyak:1981zz} in the figure.

The perturbatively calculable CF can be written as
\begin{eqnarray}
T(x,Q,\mu_{f}) &=& T^{(0)}(x,Q,\mu_{f})+a_s(\mu_r) T^{(1)}(x,Q,\mu_{f},\mu_{r}) \nonumber\\
                           &  & +a_s^{2}(\mu_r) T^{(2)}(x,Q,\mu_{f},\mu_{r}) + {\cal O}(a_s^{3}),
\label{eq3}
\end{eqnarray}
where $\mu_{r}$ is the renormalization scale and $a_s=\alpha_{s}/4\pi$. Substituting Eq.(\ref{eq2}) and Eq.(\ref{eq3}) into Eq.(\ref{eq1}), we get
\begin{eqnarray}
Q^{2} F^{\rm V}_{\pi \gamma}(Q^{2})  &=&  F^{(0)}(Q^{2})+F^{(1)}(Q^{2}, \mu_r) a_s(\mu_r)  \nonumber\\
                                     & &  +F^{(2)}(Q^{2}, \mu_r) a^{2}_s(\mu_r) + {\cal O}(a_s^{3}).
\label{eq4}
\end{eqnarray}
The LO, NLO and NNLO coefficients $F^{(i=0,1,2)}$ under the $\overline{\rm MS}$-scheme can be read from Refs.\cite{Braun:2021grd, Gao:2021iqq}. Following the standard PMC procedures, one can rewrite the pion-photon TFF as
\begin{eqnarray}
Q^{2} F^{\rm V}_{\pi \gamma}(Q^{2}) &=& 0.185+r_{1,0}a_s(\mu_r)+[r_{2,0}+\nonumber\\
&& \beta_{0}r_{2,1}]a^{2}_s(\mu_r) + {\cal O}(a_s^{3}),
\label{eq5}
\end{eqnarray}
where $\beta_{0}=11-\frac{2}{3}n_f$ with $n_{f}$ being the number of active light flavors. If setting the factorization scale as $\mu_{f}=Q$, the analytical coefficients $r_{i,j}$ are
\begin{eqnarray}
r_{1,0} &=& 0.208E_{2}(Q, \mu_0)+0.135E_{4}(Q, \mu_0)-1.230, \\
r_{2,0} &=&-7.292E_{2}(Q,\mu_0)-1.699E_{4}(Q,\mu_0)-7.015, \\
r_{2,1} &=& 2.610E_{2}(Q,\mu_0)\ln\frac{\mu^{2}_r}{Q^{2}}+0.995E_{2}(Q,\mu_0)+\nonumber\\
               & & 1.693E_{4}(Q,\mu_0)\ln\frac{\mu^{2}_r}{Q^{2}}+0.570E_{4}(Q,\mu_0)-\nonumber\\
               & & 15.461 \ln\frac{\mu^{2}_r}{Q^{2}}-2.674, \label{coe9}
\end{eqnarray}
where
\begin{eqnarray}
E_{n}(Q,\mu_0)&=& \left[\frac{\alpha_{s}(Q)}{\alpha_{s}(\mu_0)}\right]^{\gamma^{(0)}_{n}/18}.
\end{eqnarray}
Here $\gamma^{0}_{n}$ is the leading-order anomalous dimension
\begin{eqnarray}
\gamma^{0}_{n}&=&8C_{F}\left[\psi(n+2) + \gamma_{E} - \frac{3}{4}-\frac{1}{2(n+1)(n+2)}\right],
\end{eqnarray}
where $C_F=4/3$, $\gamma_{E}=0.577216$ is EulerGamma constant and $\psi(n+2)=\sum^{n+1}_{k=1}1/k-\gamma_{E}$. The function of $E_{n}(Q,\mu_0)$ is to run the moments from the initial scale $\mu_0$ to $Q$, e.g. $a_n(Q) = a_{n}(\mu_{0})E_{n}(Q,\mu_0)$. To fix the $\alpha_s$ scale running behavior, we shall adopt $\alpha_{s}(M_{Z})=0.1179\pm0.0009$~\cite{ParticleDataGroup:2022pth} to set the asymptotic scale ($\Lambda_{\rm QCD}$), which leads to $\Lambda_{\rm QCD}=0.385\pm0.015$ GeV by using the two-loop RGE.

\begin{figure}[htb]
\centering
\includegraphics[width=0.48\textwidth]{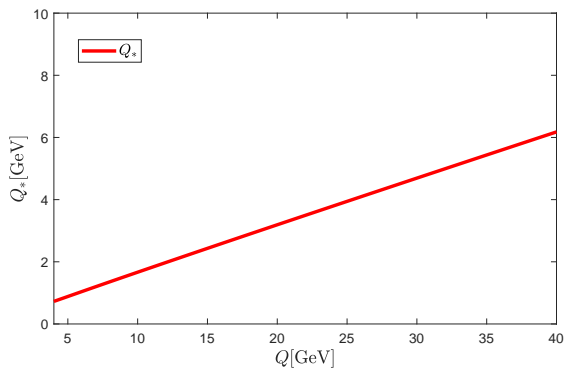}
\caption{The PMC scale $Q_{\ast}$ versus $Q$.}
\label{fig2}
\end{figure}

Following the standard PMC procedures, one can use the RG-involved $\beta_0$-terms as explicitly given in Eq.(\ref{eq5}) to fix an effective strong coupling $a_s(Q_{\ast})$, and the series becomes a conformal one independent to the choice of the renormalization scale and scheme, e.g.,
\begin{eqnarray}
Q^{2} F^{\rm V}_{\pi \gamma}(Q^{2})&=& 0.185+{r}_{1,0}a_s(Q_{\ast})+{r}_{2,0}a^{2}_s(Q_{\ast}) \nonumber\\
                                   & & + {\cal O}(a_s^{3}),
\label{eq8}
\end{eqnarray}
where the PMC scale $Q_{\ast}$ satisfies
\begin{eqnarray}
\ln\frac{Q^2_\ast}{Q^2}&=&-\frac{ r_{2,1}}{ r_{1,0}}.
\label{eq9}
\end{eqnarray}
$Q_\ast$ represents the physical momentum flow of the pion TFF in the sense of its independence to the choice of renormalization scale. Fig.(\ref{fig2}) shows the PMC scale $Q_{\ast}$ monotonous increases with the increment of $Q$.

\begin{figure}[htb]
\centering
\includegraphics[width=0.48\textwidth]{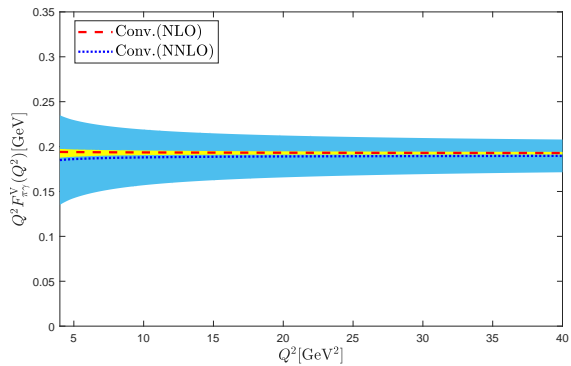}
\caption{The NLO and NNLO valence quark contribution $Q^{2} F^{\rm V}_{\pi \gamma}(Q^{2})$ under the conventional prediction, where the light-colored thinner band and the dark-colored broader band correspond to their renormalization scale uncertainties with $\mu_r\in[Q/2, 2Q]$, respectively. }
\label{fig3}
\end{figure}

\begin{figure}[htb]
\centering
\includegraphics[width=0.48\textwidth]{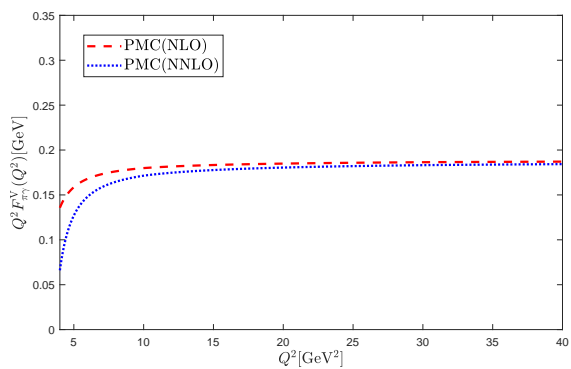}
\caption{The NLO and NNLO valence quark contribution $Q^{2} F^{\rm V}_{\pi \gamma}(Q^{2})$ under the PMC prediction.}
\label{fig4}
\end{figure}

Figs.~\ref{fig3} and \ref{fig4} present the NLO and NNLO valence quark contribution $Q^{2} F^{\rm V}_{\pi \gamma}(Q^{2})$ under the conventional and PMC predictions, respectively. Fig.\ref{fig3} shows that by including the NNLO-terms and the renormalization scale uncertainty for the conventional series of $Q^{2} F^{\rm V}_{\pi \gamma}(Q^{2})$ becomes much larger than the NLO scale uncertainty. This is caused by the fact that the large renormalization scale dependent log-terms for the conventional series, as shown by Eqs.(\ref{eq5}-\ref{coe9}), start at the NNLO level. In Fig.\ref{fig3},  as a conservation estimation, we have adopted the usual choice of $\mu_r\in[Q/2,2Q]$ to discuss the renormalization scale uncertainty. In the literature, some other choices have also been suggested to analyze the scale uncertainty. For example, the renormalization scale $\mu_r=\sqrt{\langle x \rangle} Q$ with the effective quark momentum fraction $\langle x \rangle$ within the range of $[1/4, 3/4]$ has been suggested by Ref.\cite{Gao:2021iqq}, which is based on the light-cone sum rule analysis of the pion-photon transition from factor given by Refs.\cite{Braun:1999uj, Agaev:2010aq}. Using this smaller range, a smaller NNLO scale uncertainty can be obtained, e.g. when setting $Q=10~{\rm GeV^2}$, $20~{\rm GeV^2}$ and $40~{\rm GeV^2}$, the NNLO scale uncertainty changes down from $30.65\%$ to $17.47\%$, from $23.33\%$ to $13.49\%$, and from $18.40\%$ to $10.74\%$, respectively. Fig.\ref{fig4} shows that the renormalization scale uncertainty is eliminated by applying the PMC.

Secondly, to estimate the non-valence quark contribution $Q^2 F^{\rm NV}_{\pi \gamma}(Q^{2})$, we adopt the phenomenological model suggested in Ref.\cite{Wu:2010zc}, e.g.,
\begin{eqnarray}
F^{\rm NV}_{\pi\gamma}(Q^2)&=&\frac{\alpha}{(1+\frac{Q^{2}}{\kappa^{2}})^{2}},
\end{eqnarray}
where
\begin{eqnarray}
\alpha=F^{\rm NV}_{\pi\gamma}(0)=\frac{1}{2}F_{\pi\gamma}(0)=\frac{1}{8\pi^{2}f_{\pi}},
\end{eqnarray}
and
\begin{eqnarray}
\kappa&=&\sqrt{-\frac{F_{\pi\gamma}(0)}{\frac{\partial}{\partial Q^{2}}F^{NV}_{\pi\gamma}(Q^2)|_{Q^{2}\rightarrow0}}},
\end{eqnarray}
\begin{widetext}
\begin{eqnarray}
\frac{\partial F^{NV}_{\pi\gamma}(Q^2)}{\partial Q^{2}}|_{Q^{2}\rightarrow0}&=&\frac{-A}{128\sqrt{3}m_{q}^2 \pi^2 \beta^2}\int^{1}_{0}\left(1+ b_1  C^{3/2}_{2}(2x-1)+ b_2 C^{3/2}_{4}(2x-1)\right) \frac{x}{1-x}(m_{q}^2 + 4x(1-x)\beta^2) \nonumber \\
&&  \times \exp\left(-\frac{m_{q}^2}{8\beta^2 x(1-x)}\right) dx,
\end{eqnarray}
\end{widetext}
where $m_{q}\simeq 0.3$ GeV is the constituent quark mass, and the following wave-function $\Psi_{\pi}(x,{\bf k}_{\perp})$ has been implicitly adopted to calculate the non-valence contribution
\begin{eqnarray}
\Psi_{\pi}(x,{\bf k}_{\perp})&=& A\left(1+b_1 C^{3/2}_{2}(2x-1) + b_2 C^{3/2}_{4}(2x-1)\right) \nonumber\\
&& \times \exp\left[-\frac{{\bf k}^{2}_{\perp}+m_{q}^{2}}{8\beta^{2}x(1-x)}\right], \label{wave}
\end{eqnarray}
whose dominant longitudinal behavior is given by a Gegenbauer polynomial expansion, and the parameters $A$, $b_{1,2}$ and $\beta$ are subject to the normalization condition,
\begin{eqnarray}
\int^{1}_{0}dx\int\frac{d^{2}{\bf k}_{\perp}}{16\pi^{3}}\Psi_{\pi}(x,{\bf k}_{\perp})&=&\frac{f_{\pi}}{2\sqrt{6}},
\label{eq19}
\end{eqnarray}
and the constraint from $\pi^0\to\gamma\gamma$ decay~\cite{bhl},
\begin{eqnarray}
\int^{1}_{0}dx\Psi_{\pi}(x,{\bf k}_{\perp}=0)&=&\frac{\sqrt{6}}{\pi}.
\label{eq20}
\end{eqnarray}
Here we have adopted the BHL-prescription for the transverse momentum dependence~\cite{bhl} and considered the usual helicity component~\cite{Radyushkin:2009zg}. The pion LCDA can be achieved by integrating over transverse momentum ${\bf k}_{\perp}$ from the wave-function (\ref{wave}), which can be expanded as the same form as that of Eq.(\ref{eq2}), and the first two Gegenbauer moments $a_{2}=0.206\pm0.038$ and $a_{4}=0.047\pm0.011$ will be used as two extra constraints to fix the input parameters. Using those constraints, the input parameters can be fixed numerically,
\begin{eqnarray}
A         &=& 21.22^{+0.53}_{-0.51},  \\
\beta    &=& 0.684^{+0.006}_{-0.012},  \\
b_{1} &=& 0.125^{+0.034}_{-0.028},  \\
b_{2} &=& 0.063^{+0.009}_{-0.008} ,
\end{eqnarray}
whose errors are for $\Delta a_{2}=\pm0.038$ and $\Delta a_{4}=\pm0.011$.

\begin{figure}[htb]
\centering
\includegraphics[width=0.48\textwidth]{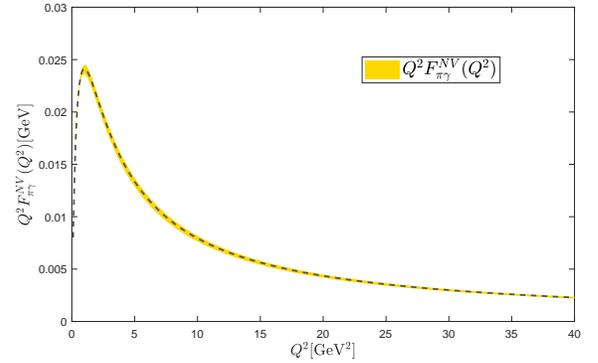}
\caption{The non-valence quark contribution $Q^2 F^{\rm NV}_{\pi \gamma}(Q^{2})$ versus $Q^{2}$. The dashed line represents its central value for $a_{2} =0.206$ and $a_{4}=0.047$, and the shaded band shows its error for $\Delta a_{2} =\pm0.038$ and $\Delta a_{4}=\pm0.011$. }
\label{fig5}
\end{figure}

Fig.(\ref{fig5}) presents the non-valence quark contribution $Q^2 F^{\rm NV}_{\pi \gamma}(Q^{2})$, whose central value corresponds to $a_{2}=0.206$ and $a_{4}=0.047$, and the error bar is for $\Delta a_{2} =\pm0.038$ and $\Delta a_{4}=\pm0.011$. It shows that because the non-valence quark contribution suffers from the $Q^2$-suppression, its magnitude is relatively small in large $Q^2$ region. In contrast, when $Q^{2}$ is small, the non-valence quark contribution becomes sizable, which should be taken into consideration to have a sound prediction for the pion-photon TFF $Q^2 F_{\pi \gamma}(Q^{2})$.

\begin{figure}[htb]
\centering
\includegraphics[width=0.48\textwidth]{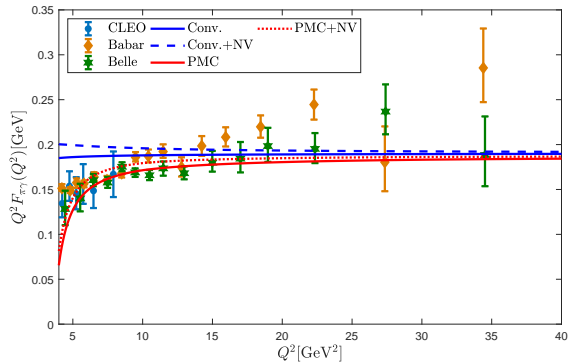}
\caption{The pion-photon TFF $Q^{2} F_{\pi \gamma}(Q^{2})$ versus $Q^2$ before and after taking into account the non-valence quark contribution. The blue-solid and red-solid lines are for the valence quark contribution under conventional and PMC predictions, respectively. The PMC prediction is scale invariant and the scale-dependent conventional prediction is for $\mu_r=Q$. The blue-dashed and the red-dotted lines are total results by taking the non-valence quark contribution into account, respectively. The experimental data of CLEO~\cite{CLEO:1997fho}, BaBar~\cite{BaBar:2009rrj}, and Belle~\cite{Belle:2012wwz} collaborations are given as a comparison. }
\label{fig6}
\end{figure}

As a combination of the NNLO-level leading-power valence quark contribution $Q^{2} F^{\rm V}_{\pi \gamma}(Q^{2})$ and the non-valence quark contribution $Q^{2} F^{\rm NV}_{\pi \gamma}(Q^{2})$, we present the total pion-photon TFF $Q^{2} F_{\pi \gamma}(Q^{2})$ in Fig.(\ref{fig6}). Here the blue-solid and red-solid lines are for the valence quark contribution $Q^{2} F^{\rm V}_{\pi \gamma}(Q^{2})$ under conventional and PMC predictions, respectively. The blue-dashed and the red-dotted lines are total results by taking the central non-valence quark contribution into account. As expected, the non-valence quark contribution is relativistically small in high $Q^2$ region, which is however sizable, especially in small $Q^2$ region. It should be emphasized that by using PMC scale-setting method, the resultant scale-invariant NNLO valence quark contribution $Q^{2} F^{\rm V}_{\pi \gamma}(Q^{2})$ becomes more convergent and improves the accuracy of the perturbation theory prediction on the valence quark effect. Numerically, when $Q^{2}\in[4,40]~{\rm GeV^{2}}$, the non-valence quark contribution to the $Q^{2} F_{\pi \gamma}(Q^{2})$ with conventional valence quark contribution $Q^{2} F^{\rm V}_{\pi \gamma}(Q^{2})$ decreases with the increment of $Q^2$ and lies within the range of $[1.21\%-8.32\%]$ for $\mu_r=Q$, which changes to $[1.24\%-23.29\%]$ for the scale-invariant PMC valence quark contribution. Specifically, when $Q^{2}=5~{\rm GeV^{2}}$, the TFF $Q^{2} F_{\pi \gamma}(Q^{2})$ increased by $7.18\%$ and $10.51\%$ for conventional and PMC valence quark results when the contribution of the non-valence quark has been taken into account. When $Q^{2}=10~{\rm GeV^{2}}$, the non-valence quark contributions become smaller and change to $4.22\%$ and $4.63\%$ for conventional and PMC valence quark predictions, respectively. When $Q^{2}>20~{\rm GeV^{2}}$, the non-valence quark contributions become less than $2\%$ for both cases.

\begin{figure}[htb]
\centering
\includegraphics[width=0.48\textwidth]{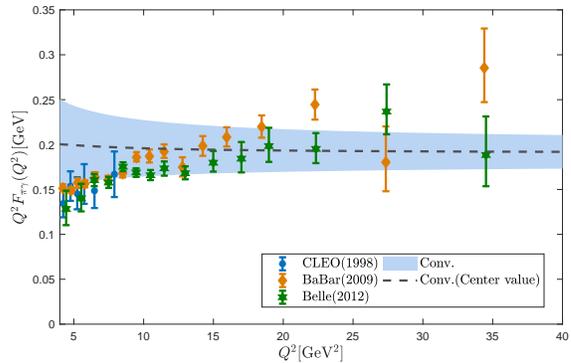}
\caption{Uncertainties of the pion-photon TFF $Q^{2} F_{\pi \gamma}(Q^{2})$ with the scale-dependent conventional valence quark contribution $Q^{2} F^{\rm V}_{\pi \gamma}(Q^{2})$. The shaded band represents the combined uncertainties caused by taking the renormalization scale $\mu_{r}\in[Q/2,2Q]$, $\Delta\alpha_{s}(M_{Z})=\pm 0.0009$, $\Delta a_{2}=\pm0.038$, $\Delta a_{4}=\pm0.011$, and the factorization scale $\mu_f\in[Q/2,2Q]$. The experimental data of CLEO~\cite{CLEO:1997fho}, BaBar~\cite{BaBar:2009rrj}, and Belle~\cite{Belle:2012wwz} collaborations are given as a comparison. }
\label{fig7}
\end{figure}

\begin{figure}[htb]
\centering
\includegraphics[width=0.48\textwidth]{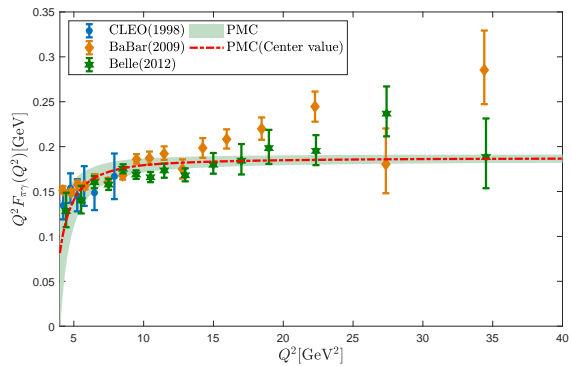}
\caption{Uncertainties of the pion-photon TFF $Q^{2} F_{\pi \gamma}(Q^{2})$ with the scale-invariant PMC valence quark contribution $Q^{2} F^{\rm V}_{\pi \gamma}(Q^{2})$. The shaded band represents the combined uncertainties caused by taking $\Delta\alpha_{s}(M_{Z})=\pm 0.0009$, $\Delta a_{2}=\pm0.038$, $\Delta a_{4}=\pm0.011$, and the factorization scale $\mu_f\in[Q/2,2Q]$. The experimental data of CLEO~\cite{CLEO:1997fho}, BaBar~\cite{BaBar:2009rrj}, and Belle~\cite{Belle:2012wwz} collaborations are given as a comparison. }
\label{fig8}
\end{figure}

Figs.(\ref{fig7}, \ref{fig8}) give the theoretical uncertainties of the pion-photon TFF $Q^{2} F_{\pi \gamma}(Q^{2})$ by taking the renormalization scale $\mu_{r}\in[Q/2,2Q]$, $\Delta\alpha_{s}(M_{Z})=\pm 0.0009$, $\Delta a_{2}=\pm0.038$, $\Delta a_{4}=\pm0.011$, and the factorization scale $\mu_f\in[Q/2,2Q]$. For the pion-photon TFF $Q^{2} F_{\pi \gamma}(Q^{2})$ with conventional valence quark contribution, its uncertainty mainly comes from renormalization scale uncertainty. For example, when $Q^{2}=20~{\rm GeV^{2}}$, the combined error is about $12.29\%$, where the renormalization scale error is $\approx 12.06\%$, the LCDA error is $\approx2.28\%$, the $\Delta\alpha_{s}(M_{Z})$ error is $\approx 0.36\%$, and the factorization scale error is $\approx0.52\%$, respectively. While, by using the PMC to eliminate the renormalization scale error of the valence quark contribution, the combined error of the TFF $Q^{2} F_{\pi \gamma}(Q^{2})$ can be significantly reduced, i.e., when $Q^{2}=20~{\rm GeV^{2}}$, the net error becomes about $2.99\%$, where the remaining LCDA error is $\approx2.89\%$ and the $\Delta\alpha_{s}(M_{Z})$ error is $\approx0.58\%$, and the factorization scale error is $\approx0.28\%$~\footnote{The elimination of the factorization scale uncertainty is a separate problem, which could be solved by matching the non-perturbative bound-state dynamics with the DGLAP evolution equation~\cite{Gribov:1972ri, Altarelli:1977zs, Dokshitzer:1977sg}. The present NNLO prediction of the pion-photon TFF shows that the factorization scale dependence can be decreased by applying the PMC for large $Q^2$ region. Two similar observations for the hadronic production of top-quark pair and Higgs boson have also been found in Refs.\cite{Wang:2014sua, Wang:2016wgw}. Then, as a practical treatment, one may apply the PMC to reduce the factorization scale uncertainty.}.

\begin{table}[htb]
\begin{center}
\begin{tabular}{  c c c  c  c c c c c }
\hline
& ~$\chi^{2}_{d.o.f}$~          & ~$\rm{CLEO}$~        & ~$\rm{Babar}$~    & ~$\rm{Belle}$~            \\
\hline
& $\rm Conventional$  & 3.81  &   23.07  & 6.88    \\
& $\rm PMC$  &   0.53    &  6.97     &  0.87     \\
\hline
\end{tabular}
\caption{The values of $\chi^{2}_{d.o.f}$ for $Q^{2} F_{\pi \gamma}(Q^{2})$ obtained with conventional and PMC prediction of $Q^{2} F^{\rm V}_{\pi \gamma}(Q^{2})$ in comparison to the CLEO, Babar, and Belle data, respectively. }
\label{tab1}
\end{center}
\end{table}

We adopt the quality of fit by using the parameter of $\chi^{2}_{d.o.f}/d.o.f$ (where the symbol ``d.o.f" is the short notation of the degree of freedom) to show to what degree the predicted pion-photon TFF $Q^{2} F_{\pi \gamma}(Q^{2})$ agrees with the data, e.g.,~\cite{ParticleDataGroup:2022pth},
\begin{equation}
\chi^{2}_{d.o.f} = \frac{1}{N}\sum_{j=1}^{N} \left[\frac{Q^{2}F(Q^{2})|_{\rm expt.}-Q^{2} F(Q^{2})|_{\rm the.}}{\sigma^{2}_{i}}\right]^{2},
\end{equation}
where ``expt" stands for the experimental value and $\sigma_{i}$ is the corresponding error for each point. The symbol of ``the" stands for the central value of the theoretical prediction. From Refs.\cite{CLEO:1997fho, BaBar:2009rrj, Belle:2012wwz}, we get $N=6$, $N=17$ and $N=15$ for CLEO, Babar and Belle data, respectively, which stands for the number of the data points. Table.\ref{tab1} shows the values of $\chi^{2}_{d.o.f}$ obtained with conventional and PMC prediction of $Q^{2} F_{\pi \gamma}(Q^{2})$ in comparison to the CLEO, Babar, and Belle data, respectively. The results indicate that the TFF $Q^{2} F_{\pi \gamma}(Q^{2})$ with the PMC valence quark contribution has much smaller $\chi^{2}_{d.o.f}$ compared to the conventional one. The conventional TFF $Q^{2} F_{\pi \gamma}(Q^{2})$ does not agree with the data, all of its corresponding $p$-values are less than $1\%$. While one may observe that by using the PMC to improve the perturbative contribution, a better prediction can be achieved. More explicitly, the PMC TFF $Q^{2} F_{\pi \gamma}(Q^{2})$ agrees with the CLEO data in the low $Q^2$ region, which corresponds to a $p$-value $\in [68\%-90\%]$. And in high $Q^2$ region, the PMC TFF $Q^{2} F_{\pi \gamma}(Q^{2})$ is close to the Belle data, which corresponds to a $p$-value $\in [50\%-68\%]$; on the contrary the PMC prediction TFF $Q^{2} F_{\pi \gamma}(Q^{2})$ is inconsistent with the Babar data, whose $p$-value is less than $1\%$.

As a summary, we have adopted the PMC single-scale approach to deal with the pQCD calculable valence quark contribution to the pion-photon TFF $Q^{2} F_{\pi \gamma}(Q^{2})$ up to NNLO accuracy. As for the TFF under the conventional scale-setting approach, as shown by Fig.(\ref{fig7}), its error is dominated by the choice of the renormalization scale. And it cannot explain the data even by including the non-valence quark contribution.

Applying the PMC, a more precise pQCD prediction for the pion-photon TFF in the large $Q^2$ region without the conventional renormalization scale ambiguity can be achieved. And as shown by Fig.(\ref{fig8}), the net theoretical uncertainty can be greatly suppressed. Using the asymptotic-like pion LCDA, we also estimate the non-valence quark contribution $Q^{2} F^{\rm NV}_{\pi \gamma}(Q^{2})$, which is sizable in the low $Q^2$ region and affects the pion-photon TFF by about $23\%$ for $Q^{2}=4~{\rm GeV^{2}}$. The resultant PMC pion-photon TFF shows a better agreement with the previous Belle data. It is hoped that previous discrepancies between the experimental measurements and theoretical predictions could be clarified by the forthcoming more precise data at the Belle II experiment~\cite{Belle-II:2018jsg}.

\hspace{2cm}

\noindent {\bf Acknowledgments:} This work was supported by the Natural Science Foundation of China under Grant No.12175025, No.12147102, No.12305091. The graduate research and innovation foundation of Chongqing, China under Grant No.CYB23011,  and by the Research Fund for the Doctoral Program of the Southwest University of Science and Technology under Contract No.23zx7122.


\begin{thebibliography}{99}

\bibitem{Adler:1969gk}
S.~L.~Adler,
``Axial vector vertex in spinor electrodynamics,''
Phys. Rev. \textbf{177}, 2426 (1969).

\bibitem{Bardeen:1969md}
W.~A.~Bardeen,
``Anomalous Ward identities in spinor field theories,''
Phys. Rev. \textbf{184}, 1848 (1969).

\bibitem{Cornwall:1966zz}
J.~M.~Cornwall,
``Current-Commutator Constraints on Three- and Four-Point Functions,''
Phys. Rev. Lett. \textbf{16}, 1174 (1966).

\bibitem{Brodsky:1971ud}
S.~J.~Brodsky, T.~Kinoshita and H.~Terazawa,
``Two Photon Mechanism of Particle Production by High-Energy Colliding Beams,''
Phys. Rev. D \textbf{4}, 1532 (1971).

\bibitem{Lepage:1979zb}
G.~P.~Lepage and S.~J.~Brodsky,
``Exclusive Processes in Quantum Chromodynamics: Evolution Equations for Hadronic Wave Functions and the Form-Factors of Mesons,''
Phys. Lett. B \textbf{87}, 359 (1979).

\bibitem{Lepage:1980fj}
G.~P.~Lepage and S.~J.~Brodsky,
``Exclusive Processes in Perturbative Quantum Chromodynamics,''
Phys. Rev. D \textbf{22}, 2157 (1980).

\bibitem{Bjorken:1968dy}
J.~D.~Bjorken,
``Asymptotic Sum Rules at Infinite Momentum,''
Phys. Rev. \textbf{179}, 1547 (1969).

\bibitem{Bjorken:1969ja}
J.~D.~Bjorken and E.~A.~Paschos,
``Inelastic Electron Proton and gamma Proton Scattering, and the Structure of the Nucleon,''
Phys. Rev. \textbf{185}, 1975 (1969).

\bibitem{Muller:1994ses}
D.~M\"uller, D.~Robaschik, B.~Geyer, F.~M.~Dittes and J.~Ho\v{r}ej\v{s}i,
``Wave functions, evolution equations and evolution kernels from light ray operators of QCD,''
Fortsch. Phys. \textbf{42}, 101 (1994).

\bibitem{Radyushkin:1996nd}
A.~V.~Radyushkin,
``Scaling limit of deeply virtual Compton scattering,''
Phys. Lett. B \textbf{380}, 417 (1996).

\bibitem{Colangelo:2014dfa}
G.~Colangelo, M.~Hoferichter, M.~Procura and P.~Stoffer,
``Dispersive approach to hadronic light-by-light scattering,''
JHEP \textbf{09}, 091 (2014).

\bibitem{Hoferichter:2018kwz}
M.~Hoferichter, B.~L.~Hoid, B.~Kubis, S.~Leupold and S.~P.~Schneider,
``Dispersion relation for hadronic light-by-light scattering: pion pole,''
JHEP \textbf{10}, 141 (2018).

\bibitem{Hoferichter:2018dmo}
M.~Hoferichter, B.~L.~Hoid, B.~Kubis, S.~Leupold and S.~P.~Schneider,
``Pion-pole contribution to hadronic light-by-light scattering in the anomalous magnetic moment of the muon,''
Phys. Rev. Lett. \textbf{121}, 112002 (2018).

\bibitem{CELLO:1990klc}
H.~J.~Behrend \textit{et al.} [CELLO],
``A Measurement of the pi0, eta and eta-prime electromagnetic form-factors,''
Z. Phys. C \textbf{49}, 401 (1991).

\bibitem{CLEO:1997fho}
J.~Gronberg \textit{et al.} [CLEO],
``Measurements of the meson - photon transition form-factors of light pseudoscalar mesons at large momentum transfer,''
Phys. Rev. D \textbf{57}, 33 (1998).

\bibitem{BaBar:2009rrj}
B.~Aubert \textit{et al.} [BaBar],
``Measurement of the gamma gamma* ---\ensuremath{>} pi0 transition form factor,''
Phys. Rev. D \textbf{80}, 052002 (2009).

\bibitem{Belle:2012wwz}
S.~Uehara \textit{et al.} [Belle],
``Measurement of $\gamma \gamma^* \to \pi^0$ transition form factor at Belle,''
Phys. Rev. D \textbf{86}, 092007 (2012).

\bibitem{Belle-II:2018jsg}
E.~Kou \textit{et al.} [Belle-II],
``The Belle II Physics Book,''
PTEP \textbf{2019}, 123C01 (2019)
[erratum: PTEP \textbf{2020}, 029201 (2020)].

\bibitem{Li:1992nu}
H.~N.~Li and G.~F.~Sterman,
``The Perturbative pion form-factor with Sudakov suppression,''
Nucl. Phys. B \textbf{381}, 129 (1992).

\bibitem{Musatov:1997pu}
I.~V.~Musatov and A.~V.~Radyushkin,
``Transverse momentum and Sudakov effects in exclusive QCD processes: Gamma* gamma pi0 form-factor,''
Phys. Rev. D \textbf{56}, 2713 (1997).

\bibitem{Li:2013xna}
H.~N.~Li, Y.~L.~Shen and Y.~M.~Wang,
``Joint resummation for pion wave function and pion transition form factor,''
JHEP \textbf{01}, 004 (2014).

\bibitem{delAguila:1981nk}
F.~del Aguila and M.~K.~Chase,
``Higher Order QCD Corrections to Exclusive Two Photon Processes,''
Nucl. Phys. B \textbf{193}, 517 (1981).

\bibitem{Braaten:1982yp}
E.~Braaten,
``QCD Corrections to Meson-Photon Transition Form-Factors,''
Phys. Rev. D \textbf{28}, 524 (1983).

\bibitem{Kadantseva:1985kb}
E.~P.~Kadantseva, S.~V.~Mikhailov and A.~V.~Radyushkin,
``Total $\alpha_{s}$ Corrections to Processes $\gamma^* \gamma^* \to \pi^0$ and $\gamma^* \pi \to \pi$ in a Perturbative {QCD},''
Yad. Fiz. \textbf{44}, 507 (1986).

\bibitem{Melic:2001wb}
B.~Melic, B.~Nizic and K.~Passek,
``BLM scale setting for the pion transition form-factor,''
Phys. Rev. D \textbf{65}, 053020 (2002).

\bibitem{Melic:2002ij}
B.~Melic, D.~Mueller and K.~Passek-Kumericki,
``Next-to-next-to-leading prediction for the photon to pion transition form-factor,''
Phys. Rev. D \textbf{68}, 014013 (2003).

\bibitem{Braun:2021grd}
V.~M.~Braun, A.~N.~Manashov, S.~Moch and J.~Schoenleber,
``Axial-vector contributions in two-photon reactions: Pion transition form factor and deeply-virtual Compton scattering at NNLO in QCD,''
Phys. Rev. D \textbf{104}, 094007 (2021).

\bibitem{Gao:2021iqq}
J.~Gao, T.~Huber, Y.~Ji and Y.~M.~Wang,
``Next-to-Next-to-Leading-Order QCD Prediction for the Photon-Pion Form Factor,''
Phys. Rev. Lett. \textbf{128}, 6 (2022).

\bibitem{Wu:2013ei}
  X.~G.~Wu, S.~J.~Brodsky and M.~Mojaza,
  ``The Renormalization Scale-Setting Problem in QCD,''
  Prog.\ Part.\ Nucl.\ Phys.\  {\bf 72}, 44 (2013).

\bibitem{Wu:2014iba}
  X.~G.~Wu, Y.~Ma, S.~Q.~Wang, H.~B.~Fu, H.~H.~Ma, S.~J.~Brodsky and M.~Mojaza,
  ``Renormalization Group Invariance and Optimal QCD Renormalization Scale-Setting,''
  Rept.\ Prog.\ Phys.\  {\bf 78}, 126201 (2015).

\bibitem{Wu:2019mky}
  X.~G.~Wu, J.~M.~Shen, B.~L.~Du, X.~D.~Huang, S.~Q.~Wang and S.~J.~Brodsky,
  ``The QCD renormalization group equation and the elimination of fixed-order scheme-and-scale ambiguities using the principle of maximum conformality,''
  Prog.\ Part.\ Nucl.\ Phys.\  {\bf 108}, 103706 (2019).

\bibitem{Brodsky:2011ta}
S.~J.~Brodsky and X.~G.~Wu,
``Scale Setting Using the Extended Renormalization Group and the Principle of Maximum Conformality: the QCD Coupling Constant at Four Loops,''
Phys.\ Rev.\ D {\bf 85}, 034038 (2012).

\bibitem{Brodsky:2011ig}
  S.~J.~Brodsky and L.~Di Giustino,
  ``Setting the Renormalization Scale in QCD: The Principle of Maximum Conformality,''
  Phys.\ Rev.\ D {\bf 86}, 085026 (2012).

\bibitem{Brodsky:2012rj}
  S.~J.~Brodsky and X.~G.~Wu,
  ``Eliminating the Renormalization Scale Ambiguity for Top-Pair Production Using the Principle of Maximum Conformality,''
  Phys.\ Rev.\ Lett.\  {\bf 109}, 042002 (2012).

\bibitem{Brodsky:2013vpa}
  S.~J.~Brodsky, M.~Mojaza and X.~G.~Wu,
  ``Systematic Scale-Setting to All Orders: The Principle of Maximum Conformality and Commensurate Scale Relations,''
  Phys.\ Rev.\ D {\bf 89}, 014027 (2014).

\bibitem{Mojaza:2012mf}
  M.~Mojaza, S.~J.~Brodsky and X.~G.~Wu,
  ``Systematic All-Orders Method to Eliminate Renormalization-Scale and Scheme Ambiguities in Perturbative QCD,''
  Phys.\ Rev.\ Lett.\  {\bf 110}, 192001 (2013).

\bibitem{Wu:2018cmb}
 X.~G.~Wu, J.~M.~Shen, B.~L.~Du, and S.~J.~Brodsky,
 Novel demonstration of the renormalization group invariance of the fixed-order predictions using the principle of maximum conformality and the $C$-scheme coupling,
 Phys.\ Rev.\ D {\bf 97}, 094030 (2018).

\bibitem{Shen:2017pdu}
 J.~M.~Shen, X.~G.~Wu, B.~L.~Du and S.~J.~Brodsky,
 ``Novel All-Orders Single-Scale Approach to QCD Renormalization Scale-Setting,''
 Phys. Rev. D \textbf{95}, 094006 (2017).

\bibitem{Zheng:2013uja}
  X.~C.~Zheng, X.~G.~Wu, S.~Q.~Wang, J.~M.~Shen, and Q.~L.~Zhang,
  ``Reanalysis of the BFKL Pomeron at the next-to-leading logarithmic accuracy,''
  J. High Energy Phys.  {\bf 10}, 117 (2013).

\bibitem{Wu:2010zc}
X.~G.~Wu and T.~Huang,
``An Implication on the Pion Distribution Amplitude from the Pion-Photon Transition Form Factor with the New BABAR Data,''
Phys. Rev. D \textbf{82}, 034024 (2010).

\bibitem{Huang:2013yya}
T.~Huang, T.~Zhong and X.~G.~Wu,
``Determination of the pion distribution amplitude,''
Phys. Rev. D \textbf{88}, 034013 (2013).

\bibitem{Radyushkin:1995pj}
A.~V.~Radyushkin,
``Quark - hadron duality and intrinsic transverse momentum,''
Acta Phys. Polon. B \textbf{26}, 2067 (1995).

\bibitem{Huang:2006wt}
T.~Huang and X.~G.~Wu,
``A Comprehensive Analysis on the Pion-Photon Transition Form Factor Involving the Transverse Momentum Corrections,''
Int. J. Mod. Phys. A \textbf{22}, 3065 (2007).

\bibitem{ParticleDataGroup:2022pth}
R.~L.~Workman \textit{et al.} [Particle Data Group],
``Review of Particle Physics,''
PTEP \textbf{2022}, 083C01 (2022).

\bibitem{Efremov:1979qk}
A.~V.~Efremov and A.~V.~Radyushkin,
``Factorization and Asymptotical Behavior of Pion Form-Factor in QCD,''
Phys. Lett. B \textbf{94}, 245 (1980).

\bibitem{Brodsky:2007hb}
S.~J.~Brodsky and G.~F.~de Teramond,
``Light-Front Dynamics and AdS/QCD Correspondence: The Pion Form Factor in the Space- and Time-Like Regions,''
Phys. Rev. D \textbf{77}, 056007 (2008).

\bibitem{Chang:2013pq}
L.~Chang, I.~C.~Cloet, J.~J.~Cobos-Martinez, C.~D.~Roberts, S.~M.~Schmidt and P.~C.~Tandy,
``Imaging dynamical chiral symmetry breaking: pion wave function on the light front,''
Phys. Rev. Lett. \textbf{110}, 132001 (2013).

\bibitem{Raya:2015gva}
K.~Raya, L.~Chang, A.~Bashir, J.~J.~Cobos-Martinez, L.~X.~Guti\'errez-Guerrero, C.~D.~Roberts and P.~C.~Tandy,
``Structure of the neutral pion and its electromagnetic transition form factor,''
Phys. Rev. D \textbf{93}, 074017 (2016).

\bibitem{Mikhailov:2016klg}
S.~V.~Mikhailov, A.~V.~Pimikov and N.~G.~Stefanis,
``Systematic estimation of theoretical uncertainties in the calculation of the pion-photon transition form factor using light-cone sum rules,''
Phys. Rev. D \textbf{93}, 114018 (2016).

\bibitem{Raya:2016yuj}
K.~Raya, M.~Ding, A.~Bashir, L.~Chang and C.~D.~Roberts,
``Partonic structure of neutral pseudoscalars via two photon transition form factors,''
Phys. Rev. D \textbf{95}, 074014 (2017).

\bibitem{RQCD:2019osh}
G.~S.~Bali \textit{et al.} [RQCD],
``Light-cone distribution amplitudes of pseudoscalar mesons from lattice QCD,''
JHEP \textbf{08}, 065 (2019).

\bibitem{Cui:2020tdf}
Z.~F.~Cui, M.~Ding, F.~Gao, K.~Raya, D.~Binosi, L.~Chang, C.~D.~Roberts, J.~Rodr\'\i{}guez-Quintero and S.~M.~Schmidt,
``Kaon and pion parton distributions,''
Eur. Phys. J. C \textbf{80}, 1064 (2020).

\bibitem{Cheng:2020vwr}
S.~Cheng, A.~Khodjamirian and A.~V.~Rusov,
``Pion light-cone distribution amplitude from the pion electromagnetic form factor,''
Phys. Rev. D \textbf{102}, 074022 (2020).

\bibitem{Stefanis:2020rnd}
N.~G.~Stefanis,
``Pion-photon transition form factor in light cone sum rules and tests of asymptotics,''
Phys. Rev. D \textbf{102}, 034022 (2020).

\bibitem{Zhong:2022lmn}
T.~Zhong, Z.~H.~Zhu and H.~B.~Fu,
``Constraints of \ensuremath{\xi}-moments computed using QCD sum rules on piondistribution amplitude models,''
Chin. Phys. C \textbf{47}, 013111 (2023).

\bibitem{Gao:2022vyh}
X.~Gao, A.~D.~Hanlon, N.~Karthik, S.~Mukherjee, P.~Petreczky, P.~Scior, S.~Syritsyn and Y.~Zhao,
``Pion distribution amplitude at the physical point using the leading-twist expansion of the quasi-distribution-amplitude matrix element,''
Phys. Rev. D \textbf{106}, 074505 (2022).

\bibitem{Zhong:2021epq}
T.~Zhong, Z.~H.~Zhu, H.~B.~Fu, X.~G.~Wu and T.~Huang,
``Improved light-cone harmonic oscillator model for the pionic leading-twist distribution amplitude,''
Phys. Rev. D \textbf{104}, 016021 (2021).

\bibitem{Chernyak:1981zz}
V.~L.~Chernyak and A.~R.~Zhitnitsky,
``Exclusive Decays of Heavy Mesons,''
Nucl. Phys. B \textbf{201}, 492 (1982)
[erratum: Nucl. Phys. B \textbf{214}, 547 (1983)].

\bibitem{Braun:1999uj}
V.~M.~Braun, A.~Khodjamirian and M.~Maul,
``Pion form-factor in QCD at intermediate momentum transfers,''
Phys. Rev. D \textbf{61}, 073004 (2000).

\bibitem{Agaev:2010aq}
S.~S.~Agaev, V.~M.~Braun, N.~Offen and F.~A.~Porkert,
``Light Cone Sum Rules for the pi0-gamma*-gamma Form Factor Revisited,''
Phys. Rev. D \textbf{83}, 054020 (2011).

\bibitem{bhl} G. P. Lepage, S.J. Brodsky,T. Huang, and P. B. Mackezie,
in Particles and Fields, Proceedings of the Banff
Summer Institute on Particle Physics, Banff, Alberta,
Canada, 1981 2, edited by A. Z. Capri and A. N. Kamal
(Plenum, New York, 1983), p. 83.

\bibitem{Radyushkin:2009zg}
A.~V.~Radyushkin,
``Shape of Pion Distribution Amplitude,''
Phys. Rev. D \textbf{80}, 094009 (2009).

\bibitem{Gribov:1972ri}
 V.~N.~Gribov and L.~N.~Lipatov,
 ``Deep inelastic e p scattering in perturbation theory,''
 Sov.\ J.\ Nucl.\ Phys.\ {\bf 15}, 438 (1972).

\bibitem{Altarelli:1977zs}
 G.~Altarelli and G.~Parisi,
 ``Asymptotic Freedom in Parton Language,''
 Nucl.\ Phys.\ B {\bf 126}, 298 (1977).

\bibitem{Dokshitzer:1977sg}
 Y.~L.~Dokshitzer,
 ``Calculation of the Structure Functions for Deep Inelastic Scattering and e+ e- Annihilation by Perturbation Theory in Quantum Chromodynamics.,''
 Sov.\ Phys.\ JETP {\bf 46}, 641 (1977).

\bibitem{Wang:2014sua}
 S.~Q.~Wang, X.~G.~Wu, Z.~G.~Si and S.~J.~Brodsky,
 ``Application of the Principle of Maximum Conformality to the Top-Quark Charge Asymmetry at the LHC,''
 Phys. Rev. D \textbf{90}, 114034 (2014).

\bibitem{Wang:2016wgw}
 S.~Q.~Wang, X.~G.~Wu, S.~J.~Brodsky and M.~Mojaza,
 ``Application of the Principle of Maximum Conformality to the Hadroproduction of the Higgs Boson at the LHC,''
 Phys. Rev. D \textbf{94}, 053003 (2016).

\end{thebibliography}
\end{document}